\documentclass[notitlepage,twocolumn,letterpaper,natbib,aps,prd,amsmath,amsfonts,nofootinbib,preprintnumbers,superscriptaddress,secnumarabic]{revtex4-1}
\pdfoutput=1

\usepackage{amssymb,amsmath,latexsym,mathrsfs}
\usepackage{physics}
\usepackage{mathtools}
\usepackage{graphicx}
\usepackage{dcolumn}
\usepackage{multirow}
\usepackage{booktabs}
\usepackage{bbold}
\usepackage{algorithm}
\usepackage{algpseudocode}
\usepackage{adjustbox}
\usepackage[shortlabels]{enumitem}
\usepackage{bm}
\usepackage[usenames,dvipsnames]{color}
\usepackage[breaklinks,colorlinks,urlcolor=magenta,citecolor=magenta,linkcolor=magenta]{hyperref}
\usepackage{xcolor}
\usepackage{booktabs}
\usepackage{tikz}
\usepackage{xspace}
\usepackage{fontawesome}
\usepackage{cancel}
\usetikzlibrary{positioning}
\usetikzlibrary{shapes.geometric}
\usetikzlibrary{decorations.pathreplacing,calligraphy}
\usetikzlibrary{arrows.meta, positioning}

\definecolor{linkcolor}{RGB}{31, 118, 180} 
\definecolor{citecolor}{RGB}{174, 60, 60} 
\definecolor{myaqua}{RGB}{51, 255, 255}
\hypersetup{
  linkcolor  = myaqua!50!black,
  citecolor  = citecolor,
  urlcolor   = myaqua!50!black,
  colorlinks = true
}

\DeclareMathOperator*{\E}{\mathbb{E}}

\newcommand{\eg}{{e.g.}\xspace}
\newcommand{\ie}{{i.e.}\xspace}

\begin{document}

\title{Tests for model misspecification in simulation-based inference: \\ from local distortions to global model checks}

\author{Noemi Anau Montel}
\email[ ]{noemiam@mpa-garching.mpg.de}
\affiliation{Max-Planck-Institut für Astrophysik, Karl-Schwarzschild-Str.\ 1, 85748 Garching, Germany}

\author{James Alvey}
\email[ ]{jbga2@cam.ac.uk}
\affiliation{Kavli Institute for Cosmology Cambridge, Madingley Road, Cambridge CB3 0HA, United Kingdom}
\affiliation{Institute of Astronomy, University of Cambridge, Madingley Road, Cambridge CB3 0HA, United Kingdom}

\author{Christoph Weniger}
\email[ ]{c.weniger@uva.nl}
\affiliation{GRAPPA Institute, Institute for Theoretical Physics Amsterdam,
University of Amsterdam, Science Park 904, 1098 XH Amsterdam, The Netherlands}

\begin{abstract}
    \noindent Model misspecification analysis strategies, such as anomaly detection, model validation, and model comparison are a key component of scientific model development. 
    Over the last few years, there has been a rapid rise in the use of simulation-based inference (SBI) techniques for Bayesian parameter estimation, applied to increasingly complex forward models. To move towards fully simulation-based analysis pipelines, however, there is an urgent need for a comprehensive simulation-based framework for model misspecification analysis. 
    In this work, we provide a solid and flexible foundation for a wide range of model discrepancy analysis tasks, using \textit{distortion-driven model misspecification tests}.
    From a theoretical perspective, we introduce the statistical framework built around performing many hypothesis tests for distortions of the simulation model. We also make explicit analytic connections to classical techniques: anomaly detection, model validation, and goodness-of-fit residual analysis. Furthermore, we introduce an efficient self-calibrating training algorithm that is useful for practitioners. We demonstrate the performance of the framework in multiple scenarios, making the connection to classical results where they are valid. Finally, we show how to conduct such a distortion-driven model misspecification test for real gravitational wave data, specifically on the event GW150914.
\end{abstract}

\maketitle

\section{Introduction} \label{sec:introduction}

\noindent The primary goal of the physical sciences is to refine analytic or computational models that form the backbone of our understanding of physical phenomena. Key steps in this process are designing, fitting, and validating different models against data.
There are a variety of existing strategies to approach this. Powerful tools include Bayesian evidence estimation conditioned on observations for model comparison~\cite{skilling_nested_2006}, or goodness-of-fit and hypothesis tests for establishing the validity of a model across the observed data and making discoveries~\cite{neyman_pearson, the_atlas_collaboration_observation_2012}. 

In recent years, simulation-based inference (SBI), also known as implicit-likelihood inference, has emerged as an important tool for inference when simulations from implicitly-defined models are available \cite{cranmer_frontier_2020}. This approach is especially useful for managing the increasing dimensionality of datasets and the growing complexity of scientific models, where often a full probabilistic description is difficult or even impossible to define. It is also useful in the regime where classical methods are computationally demanding, but efficient simulations remain feasible.

In SBI, the modeling complexity is shifted from having to define a likelihood function to having to program a simulator, which may be easier than constructing an analytical probabilistic description. In principle, this allows one to include and account for more effects in the modeling than in traditional methods.
However, as in any statistical inference framework, the question of model misspecification --- how to detect it, how it affects parameter reconstruction, and how to account for it --- remains. In the SBI setting, model misspecification occurs when the data-generating function (the simulator) is a poor representation of the physical processes being analyzed. Modeling choices in the simulator are thus extremely important and the degree of simulator complexity requires a careful balance between over-fitting and under-fitting the data.

The majority of existing SBI applications have so far focused on uncertainty quantification in parameter inference tasks \cite[\eg][]{Mishra-Sharma:2021oxe, Montel:2022fhv, Tucci:2023bag, abellan, Alsing:2019xrx, Modi:2023llw, Karchev:2024zpu, Bhardwaj:2023xph, Dax:2021tsq}.~\footnote{An extensive list of SBI applications can be found here: \url{https://github.com/smsharma/awesome-neural-sbi}.} Recently though, there has been a surge of interest in the development of SBI algorithms for other cornerstones of classical statistics: for example, model comparison through Bayesian evidence computation \cite{jeffrey_evidence_2024, Gessey-Jones:2023vuy, mancini_bayesian_2023}, and frequentist hypothesis testing \cite{cranmer2016approximating,dalmasso2020confidence, heinrich_learning_2022}. In particular, calibrated binary classifiers have been shown to be equivalent to classical likelihood-ratio test statistics \cite{cranmer2016approximating}. It has also been shown that it is possible to build confidence intervals with good frequentist properties \cite{dalmasso2020confidence}, and obtain test statistics equivalent to profile likelihood ratios \cite{heinrich_learning_2022} in SBI settings. A recent example of this interest is the simulation-based cosmological analysis of KiDS-1000 data, where a classical goodness-of-fit measure for Bayesian settings has been adapted to inspect SBI posteriors \cite{von_wietersheim-kramsta_kids-sbi_2024}.

The effects of model misspecification in SBI and possible mitigation strategies have also received some attention in the recent literature. Generally, a misspecification of the model is expected to lead to wrong inference results. In likelihood-based inference, a misspecified likelihood function is expected to always lead to the same incorrect results~\cite{white_maximum_1982}. In contrast, it has been shown that the failure modes of SBI depend on the adopted method; this means that a consistency check between methods can be used as a diagnostic check~\cite{cannon_investigating_2022}. Other diagnostic checks are based on testing whether the learned data summaries lead to outliers in latent space when applied to real-world data~\cite{schmitt_detecting_2024-1}. Lastly, in order to reduce the effect of model misspecification on inference results, various approaches to use a small set of (labeled or unlabeled) real-world data examples to make data summaries resilient against misspecification have been explored~\cite{huang_learning_2023, wehenkel_addressing_2024, ward_robust_2022, dellaporta_robust_2022}. In this work, we will limit ourselves to developing diagnostic strategies for \emph{detecting model misspecification}, which may serve as the basis for identifying and eventually correcting any insufficiency in a given simulation model.

With this context in mind, we propose a novel SBI framework for a wide variety of \emph{locally interpretable} and \emph{globally significant} model misspecification tests for simulated hypotheses. Our versatile and flexible framework is based on \emph{model augmentation}, fully embracing the core aspect of SBI: if you can simulate it, you can test for it. In general, the proposed framework allows one to perform, in a practical and comprehensive way, analyses such as anomaly detection and model validation. It also allows us to assess their significance, and perform residual analyses. In limiting cases, we demonstrate the close connection of our approach to classical matched filtering and $\chi^2$-goodness-of-fit tests. Furthermore, we propose an efficient self-calibrating training algorithm that converges to look for distortions that are just plausible given the observational noise. We demonstrate the performance of the framework in an instructive scenario, and show a proof-of-concept application of the framework to real gravitational waves data.

The rest of the paper is organized as follows. In Section \ref{sec:method}, we motivate and describe our framework based on high-volume hypothesis testing via data augmentation. Section \ref{sec:example} displays an instructive example, highlighting its direct link to traditional techniques and analytic results. We present an application to gravitational waves data in Section~\ref{sec:gw}. In Section \ref{sec:discussion}, we discuss possible improvements and the relevant limitations of our approach. Finally, we present some outlook and our conclusions in Section \ref{sec:conclusion}. 

\vspace{3pt}
\noindent \texttt{Code:} The code to reproduce the examples in this work can be found at \href{https://github.com/NoemiAM/mist}{\faGithub\,\texttt{NoemiAM/mist}}.  

\section{Misspecification testing in simulation-based inference} \label{sec:method} 

\begin{figure*}
    \centering
    \includegraphics[width=0.85\linewidth,clip]{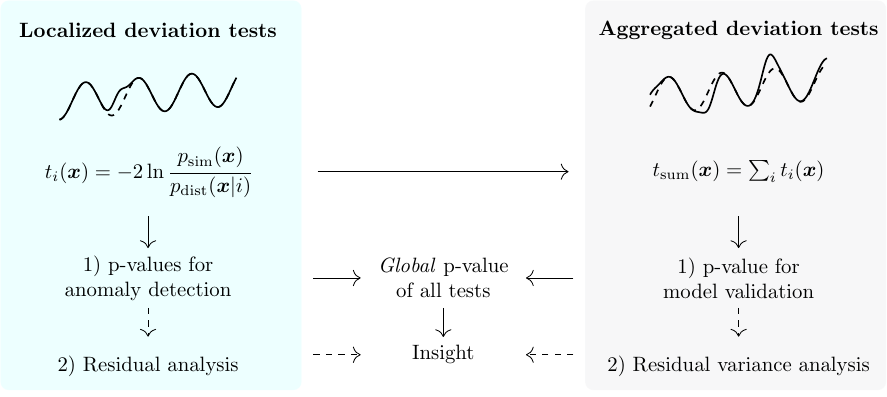}
    \caption{Summary illustration of the presented framework for tests of model misspecification in SBI (see Section~\ref{sec:method} for details). \textbf{Left panel}: An ensemble of localized test statistics is learned by neural networks (see Appendix~\ref{app:BCE} for details); they are typically more sensitive towards isolated distortions and in some limits can be the basis for anomaly detection. Their individual significance can be quantified with Monte-Carlo estimates, and in specific training scenarios (see Appendix~\ref{app:SNR} for details) one can visualize the distortions to the model in data space through residuals. \textbf{Right panel}: Aggregated test statistics can be constructed given any subset of localized test statistics; they are sensitive towards the cumulative evidence of multiple distortions and in some limits can be the basis for model validation tests. Their individual significance can be quantified with Monte-Carlo estimates, and in specific training scenarios one can perform a residual variance analysis. \textbf{Central panel}: We can estimate the overall global significance of all the performed tests, accounting for their correlation.}
    \label{fig:summary}
\end{figure*}

\noindent In general, we are interested in testing and searching for aspects of the data that might not be fully accounted for by our base model. These deviating features can arise in the form of, \eg, distortions in individual data bins, correlated distortions, excesses in specific Fourier modes, or additional model components, depending on the type of data at hand. Here, we describe a general SBI framework to \emph{simultaneously} test for many types of deviations with respect to the base model in the data and discuss its connections to classical testing frameworks.

The proposed framework is summarized in Figure~\ref{fig:summary}. It is based on a high-volume (\ie large numbers of individual tests) hypothesis testing algorithm rooted in SBI (Section~\ref{subsec:hypothesis_test}), that broadly recovers classical techniques in anomaly detection and model validation in limiting cases. Given the large number of tests performed, it is necessary to account for correlated trials when assessing the overall significance. This is discussed in Section~\ref{subsec:global}. Furthermore, we make explicit connections between the proposed framework and classical ones in Section~\ref{subsec:connection}. Finally, we discuss briefly training strategies in Section~\ref{subsec:training}.

\subsection{High-volume hypothesis testing} \label{subsec:hypothesis_test}

\noindent Our starting point for carrying out model misspecification testing using many different types of distortion (what we call \textit{high-volume}) is the classical hypothesis testing framework~\cite[e.g.][]{cranmer2015practical}. 
In a nutshell, hypothesis testing aims to assess a null hypothesis, $H_0$, by determining whether it can be rejected in favor of an alternative hypothesis, $H_1$, given observational data ${\bm x}_{\rm obs}$. In practice, this involves defining a test statistic $t(\bm{x})$, a single real-valued summary, ideally designed to maximize the ability to distinguish between $H_0$ and $H_1$. Evaluated on the observed data, the test statistic yields $t_{\rm obs} = t(\bm{x}_{\rm obs})$, which is then compared with its distribution under $H_0$, denoted as $p(t|H_0)$. The key quantity of interest in this process is the $\mathrm{p}$-value, defined as the probability of observing a value of $t$ at least as extreme as the observed one, $t_{\rm obs}$, assuming the null hypothesis is true, $\mathrm{p}_{\rm obs} = p(t > t_{\rm obs}) = \int_{t_{\rm obs}}^\infty p(t|H_0) \dd t $.

In an SBI setting, the logical first step is to define the null hypothesis in terms of the base simulator, $p_\text{sim}(\bm{x})$, that implicitly defines the likelihood of the base model.~\footnote{In general, the typical setup will be parametric simulators in the form of $p_\text{sim}(\bm{x}) = \int \dd\Theta\ p(\bm{x}|\Theta)p(\Theta)$, where $p(\bm{x}|\Theta)$ is the likelihood of the data given some model parameters $\Theta$ with prior $p(\Theta)$. One can also choose to test simulators with constrained proposal distribution through active learning~\cite{miller_truncated_2021, papamakarios2019sequentialneurallikelihoodfast, AnauMontel:2023stj}. More generally, $x \sim p_\text{sim}(\bm{x})$ can also be samples from generative models.}
The null hypothesis is thus defined as
\begin{equation}
    H_0: \bm{x} \sim 
    p_\text{sim}(\bm{x})
    \equiv 
    p(\bm{x} | H_0) \;.
\end{equation}

To test for deviations in the data that are not fully described by the base model/simulator,
we construct an \emph{ensemble} of $N_\text{alt}$ alternative data-generating functions, \ie\ alternative simulation models. 
These alternative simulation models are defined by suitably augmenting the base simulator model with distinct \emph{stochastic distortions $i$ to the data}, 
\begin{equation}
    \tilde{\bm x} \sim p_\text{dist}(\tilde{\bm x}| i)
    : \tilde {\bm x} \sim
    p_\text{dist}(\tilde{\bm x}|\bm x, i)
    \quad\text{with }
    \bm x \sim p_\text{sim}(\bm x)
\end{equation}
The stochastic distortions model any deviating feature of the data we want to test for. 
Each of these possible distortions defines an alternative hypothesis
\begin{equation}
    H_i: \bm{x}_i \sim p_\text{dist}({\bm x}| i) \equiv p(\bm{x} | H_i) \quad \text{with} \quad i = 1, \dots, N_\text{alt} \;,
\end{equation}
against which we test our base model $H_0$. In general, the index $i$ that characterizes the distortion does not have to be discrete, \ie\ one could parameterize the distortions in a continuous way.~\footnote{For example, particle physics ``bump-hunt" searches~\cite{the_atlas_collaboration_observation_2012} are one common scenario in which a continuous labeling is natural.} In this work, however, we use the discrete indexing.

As for the test statistic, we consider the widely used log-likelihood ratio test statistic\footnotemark[4]
\addtocounter{footnote}{1}
\begin{equation} \label{eq:ti}
    t_i(\bm{x}) \equiv  - 2 \ln\frac{p(\bm{x}| H_0)}{p(\bm{x} | H_i)} = -2 \ln
    \frac
    {p_\text{sim}(\bm x)}
    {p_\text{dist}(\bm x | i)} \;.
\end{equation}
In the applications below, we will approximate this ensemble of $N_\text{alt}$ test statistics $t_i$ via neural networks. The learned neural test statistics allow us to test for multiple types of deviations in the data simultaneously. We refer the reader to Section~\ref{subsec:training} for more details regarding specific training strategies. 

It is important to emphasize that, within SBI, ensemble hypothesis testing for model distortions is achieved \emph{operationally} through data augmentation. In simple terms, different hypothesis $H_i$, each describing the data distribution via a different likelihood functions $p(\bm{x} | H_i)$, manifest in SBI through distinct data generation processes --- simulation models with diverse stochastic distortions, $p_\text{dist}({\bm x}|i)$ --- which implicitly encode the alternative likelihoods. 
The flexibility of neural networks enables high-volume hypothesis testing for all these distortions.

\footnotetext[4]{We know from the Neyman-Pearson lemma \cite{neyman_pearson} that the log-likelihood ratio provides the optimal test statistic to maximally distinguish between $H_0$ and $H_1$ in case of simple hypotheses (\ie hypotheses that fully specify the probability distribution of the data). For composite hypotheses (\ie hypotheses where some of the distribution parameters are not specified), the log-likelihood ratio test is usually generalized by optimizing the likelihood over the parameter spaces of both the null and alternative hypotheses \cite{Wilks:1938}. We note that in most of the use-cases this framework can be applied to, one usually does not have a uniformly most powerful test.}

\subsection{Localized and aggregated tests} \label{subsec:aggregated}

\noindent We refer to each test statistic $t_i$ as \emph{localized}, not in the spatial sense, but because it is tied to a specific, single, narrowly defined distortion scenario in the data. 
These localized test statistics are more sensitive towards single isolated distortions, and, in some limits, lead to matched filter and anomaly localization ``bump-hunt" type of analyses.

For any subset of alternative hypotheses $H_i$, we can also consider the sum over all localized test statistics,
\begin{equation}\label{eq:tsum}
    t_\mathrm{sum}(\bm x) = \sum_{i=0}^{N_\mathrm{alt}}  t_i(\bm{x})\;,
\end{equation}
which is expected to be particularly useful if there is a general tendency to prefer alternative hypothesis $H_i$ over $H_0$. This \emph{aggregated} test statistic provides \emph{complementary} information about the statistical significance of favoring the alternatives $H_i$ over the baseline model $H_0$. It is thus sensitive towards the cumulative evidence of multiple distortions being present in the data. The precise meaning of $ t_\mathrm{sum}(\bm x)$ depends on the specific choice of the subset of alternative hypotheses $H_i$, allowing flexibility in designing custom tests that target particular classes of deviations (for a short discussion see Section~\ref{sec:discussion}). For example, we will see that under certain conditions this aggregated test statistic asymptotically follows a $\chi^2$ distribution. The connection of the framework to classical tests will be discussed in Section~\ref{subsec:connection}, with more details provided in Appendix~\ref{app:math}.

\subsection{Individual and global significance estimates} \label{subsec:global}

\noindent A sufficiently large test statistic for a given discrepancy (whether a localized one $t_i$ or an aggregated one $t_\mathrm{sum}$) hints that the corresponding alternative hypothesis is preferred over the baseline assumption $H_0$. As is widely done, to correctly quantify the statistical significance of a discrepancy, one can use a Monte Carlo estimate of the corresponding $\mathrm{p}$-value.\footnote{
Operationally, this is easily computed as 
\begin{equation}\label{eq:p-local}
    {\text{p}}(\bm x_\text{obs}) = \mathbb{E}_{\bm x \sim p_\text{sim}(\bm{x})}
    \left[{\Theta}\left(
    {t}(\bm{x})
    -{t}(\bm{x}_\text{obs})
    \right)
    \right]\;,
\end{equation}
where $\Theta[\cdot]$ is the Heaviside step function; or equivalently 
\begin{equation}\label{eq:p-local}
    {\text{p}}(\bm x_\text{obs}) = \mathbb{E}_{\bm x \sim p_\text{sim}(\bm{x})}\ \mathbb{I}
    \left[(
    {t}(\bm{x})
    >{t}(\bm{x}_\text{obs}) 
    \right]\;,
\end{equation}
where $\mathbb{I}[\cdot]$ is an indicator function, which is unity if the condition is true and zero otherwise.\label{footnote:mc}
}
This estimate requires a sufficiently large number of samples from the base model, but can be done in parallel for all alternative hypotheses. 

The above construction provides information about the \emph{individual} significance of each specific alternative (\eg\ a localized anomaly or a cumulative effect of many distortions), without accounting for the fact that multiple hypotheses (\eg\ the presence of a specific localized distortion or the possibility that there are many subtle correlated discrepancies not being modeled) are being tested.
Therefore, we need to account for the fact that, when testing a large number of alternative hypotheses, the probability of observing a large test statistic purely by chance increases.
This is analogous to the ``look-elsewhere effect" in particle physics \cite{Gross_2010}, where searching over many possible signal locations increases the chance of a statistical fluctuation appearing significant.

To estimate the significance of the individual discrepancies from the null hypothesis $H_0$ at a more \emph{global} level, accounting for the large number of performed tests and their possible correlations, it is necessary to estimate a global $\mathrm{p}$-value for the overall, \ie\ trials-corrected, significance of the minimum observed $\mathrm{p}$-value from all tests. In different words, the global $\mathrm{p}$-value estimates the probability of observing the most extreme test outcome across any of the hypotheses being tested.

Operationally, this global $\mathrm{p}$-value can be obtained as follows. First, for $N_{\mathrm{mc}}$ Monte Carlo samples, we compute p-values (see Footnote~\ref{footnote:mc}) for all of the $N_\mathrm{t}$ individual tests (either localized or aggregated) of interest. Each of these Monte Carlo samples generates $N_\mathrm{t}$ p-values, one for each test. While any single p-value is uniformly distributed under the null hypothesis, it is important to note that they exhibit correlations among themselves. To account for this, from each of the $N_{\mathrm{mc}}$ samples, we extract the minimum p-value across the $N_\mathrm{t}$ tests, yielding $N_{\mathrm{mc}}$ such minima. Now, we want to ask the following question: given an observation on which we have performed $N_\mathrm{t}$ tests, what is the probability of having observed the most significant deviation (\ie the one with the test yielding the minimum p-value)? We can use the $N_{\mathrm{mc}}$ minimum p-values to answer this question by comparing the distribution of these minimum p-values with the minimum p-value of the observation. This allows us to properly assess the global significance of the observed minimum p-value, the global $\mathrm{p}$-value.

\subsection{Connection to classical testing frameworks} \label{subsec:connection}

\noindent Under a few key assumptions, the above framework for misspecification testing in SBI via high-volume hypothesis testing can be connected to classical testing methods. For a detailed derivation of the following results we refer the reader to Appendix~\ref{app:math}.

Given a base simulator from which one can sample $\bm{x} \sim p(\bm{x}| H_0)$, let us consider simple \emph{stochastic additive non-Gaussian distortions} with specific noise directions $\bm{n}^{(i)}$ in data space
\begin{equation} \label{eq:additive1}
    H_i: 
    \tilde{\bm{x}} = \bm{x} + \epsilon \cdot \bm{n}^{(i)} \quad \text{with} \quad \epsilon \sim \mathcal{U}(-b, b)\;.
\end{equation}

Assuming a Gaussian likelihood function for the base model, in the large sample limit, and for scenarios where the maximum-likelihood estimator is not significantly correlated with the distortion $\bm{n}^{(i)}$, we derive the central result that \textbf{the test statistic for a given distortion $\bm{n}^{(i)}$ is directly related to the signal-to-noise ratio (SNR) of that distortion in the data}
\begin{equation} \label{eq:t_snr}
    t_i(\bm{x}) \simeq \text{SNR}_i^2(\bm{x}) + C
\end{equation}
where $C$ is a constant that depends on the prior distribution of the distortion amplitude $\epsilon$ and is defined in Equation~\ref{eq:c}, and the $\text{SNR}$ is
\begin{equation} \label{eq:snr}
    \text{SNR}_i(\bm x) 
    = \frac{\epsilon^\ast(\bm x)}{\sigma}\;,
\end{equation}
where $\epsilon^\ast(\bm x)$ is the maximum likelihood estimator (MLE) of $\epsilon$ (Equation~\ref{eq:epsilon_gauss}), and $\sigma^2$ the variance of the MLE (Equation~\ref{eq:sigma_gauss}).

This is directly analogous to the matched filtering technique used in signal detection, where the data is correlated with a set of template signals to find the one with the highest SNR. Performing high-volume hypothesis testing in this limiting case and under the above assumptions, allows one to obtain results simultaneously for many types of distortions $\bm{n}^{(i)}$. 

Let us now restrict the generality of the distortions such that each distortion corresponds to a deviation along one of the \emph{standard basis} vectors in data space, \ie\ $\bm n^{(i)} = \bm e_i$, where $\bm e_i$ is the unit vector in the $i\-\text{th}$ data space dimension. Furthermore, let us consider a diagonal noise covariance matrix. Under the same assumptions as before, we find that the sum of the test statistics over all alternative hypotheses $H_i$ for the stochastic additive distortions under consideration, \ie\ the aggregated test statistic from Equation~\ref{eq:tsum}, is related to the Pearson's $\chi^2$ test statistic (see derivation for Equation~\ref{eq:app_chi2}):
\begin{equation} 
    t_\text{sum}(\bm{x}) = \chi^2(\bm{x}) + \mathrm{const} \;. 
\end{equation}
Thus, the sum test statistic captures the cumulative effect of small deviations across multiple dimensions, reflecting an overall mismatch between the data and the model, and providing a goodness-of-fit measure.

\subsection{Training strategies} \label{subsec:training}

\noindent In this work, we adopt two different training strategies to estimate the localized test statistics $t_i$. The first one is very general and does not rely on any prior assumptions, whereas the second one is motivated by the connection of our framework to classical testing frameworks as discussed in the previous section.

\vspace{5pt}
\noindent \textbf{BCE -} The first training strategy employs discriminative classifiers to approximate likelihood ratio statistics using the binary cross entropy (BCE) loss \cite{cranmer2016approximating}. Thus, we refer to it as the {BCE} training strategy. For more details we refer the reader to Appendix~\ref{app:BCE}.

\vspace{5pt}
\noindent \textbf{SNR -} The second training strategy stems from the results discussed in Section~\ref{subsec:connection}, where it was shown that under certain assumptions the test statistic for a given distortion $\bm{n}^{(i)}$ is directly related to the SNR of that distortion in the data (Equation~\ref{eq:t_snr}). Hence, these test statistics can be equivalently trained by minimizing a Gaussian negative log-likelihood loss for the MLE of the matched filter $\epsilon^\ast(\bm x)$ and its variance $\sigma^2$. Thus, we refer to it as the {SNR} training strategy. For more details we refer the reader to Appendix~\ref{app:SNR}.

\vspace{5pt}
\noindent A direct comparison of the training strategies for the example presented in Section~\ref{sec:example} can be found in Appendix~\ref{app:comparison}. We also apply both training strategies to the gravitational waves example in Section~\ref{sec:gw}.

\section{Instructive example} \label{sec:example}

\begin{figure*}
    \centering
    \begin{tikzpicture}
    \node at (0cm,0cm) {\includegraphics[height=12cm,clip]{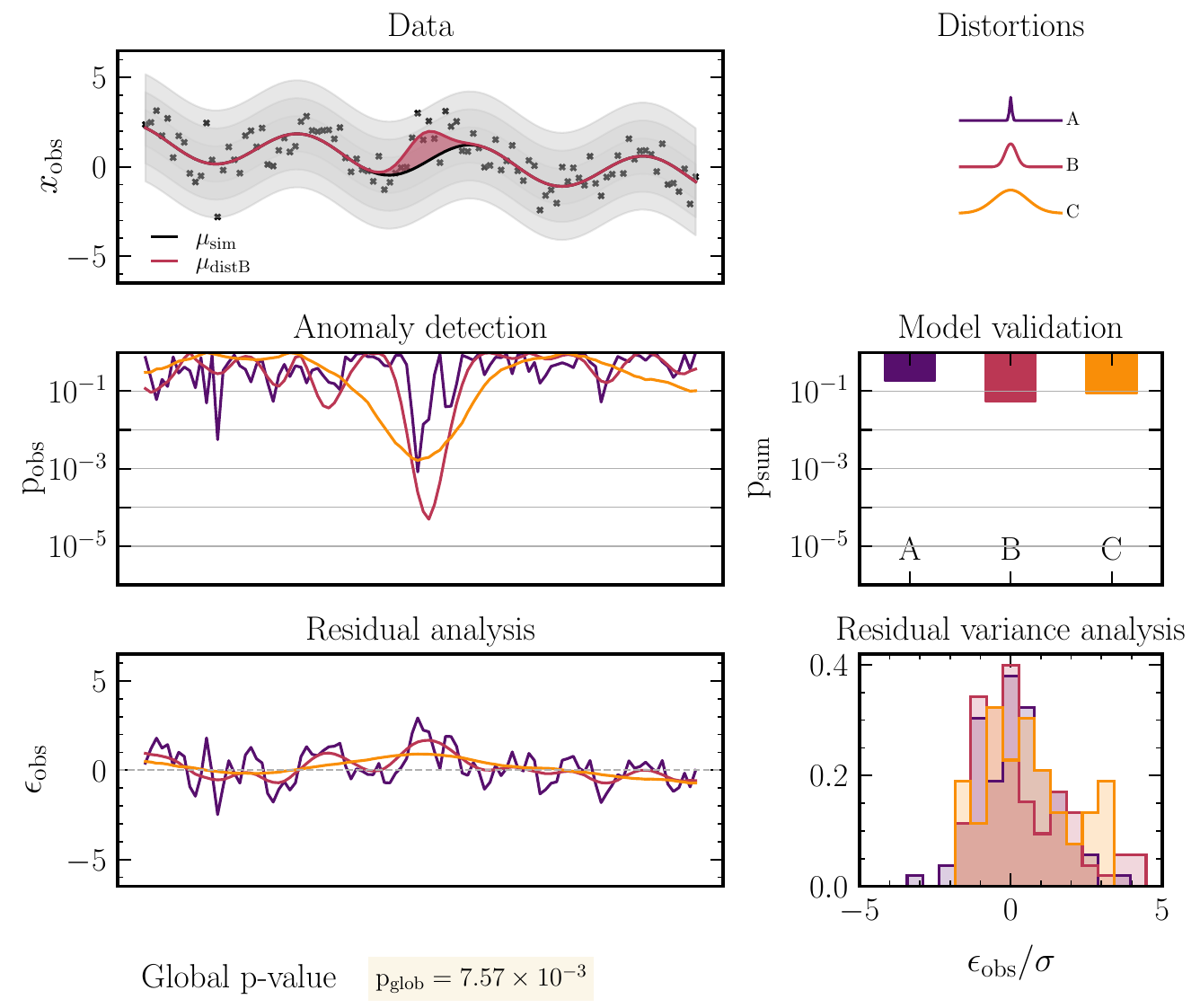}};
    \end{tikzpicture}
    \caption{Comprehensive summary of the framework results for the instructive example presented in Section~\ref{sec:example}. The panels show the results by following the structure of our framework, as represented in the summary graphic Figure~\ref{fig:summary}.
    The \textbf{upper-left panel} depicts scattered data points $\bm{x}_\mathrm{obs}$, the baseline signal $\bm{\mu}_\mathrm{sim}$, and the signal distorted by an additive stochastic distortion of type B, $\bm{\mu}_\mathrm{distB}$, as described in Section~\ref{sec:example}. The gray bands highlights the 1-, 2-, and 3-$\sigma$ regions of the baseline Gaussian noise.
    The \textbf{upper-right panel} visualizes the three types of deviations with different correlation scale under investigation. The results from different networks is color-coded based on the type of distortion they were trained on in the following panels. 
    The \textbf{center-left panel} showcases the significance of localized distortions, the \textbf{center-right panel} the significance of the aggregated distortions, and the \textbf{bottom text} the global significance.
    Finally, the \textbf{bottom panels} the strength of the distortions in data space and their variance. These latter results are achievable only through the SNR training strategy (see Section~\ref{subsec:training}).
    }
    \label{fig:plot1}\vspace*{-6pt}
\end{figure*}

\begin{figure*}
    \centering
    \begin{tikzpicture}
    \node at (0cm,0cm) {\includegraphics[height=12cm,clip]{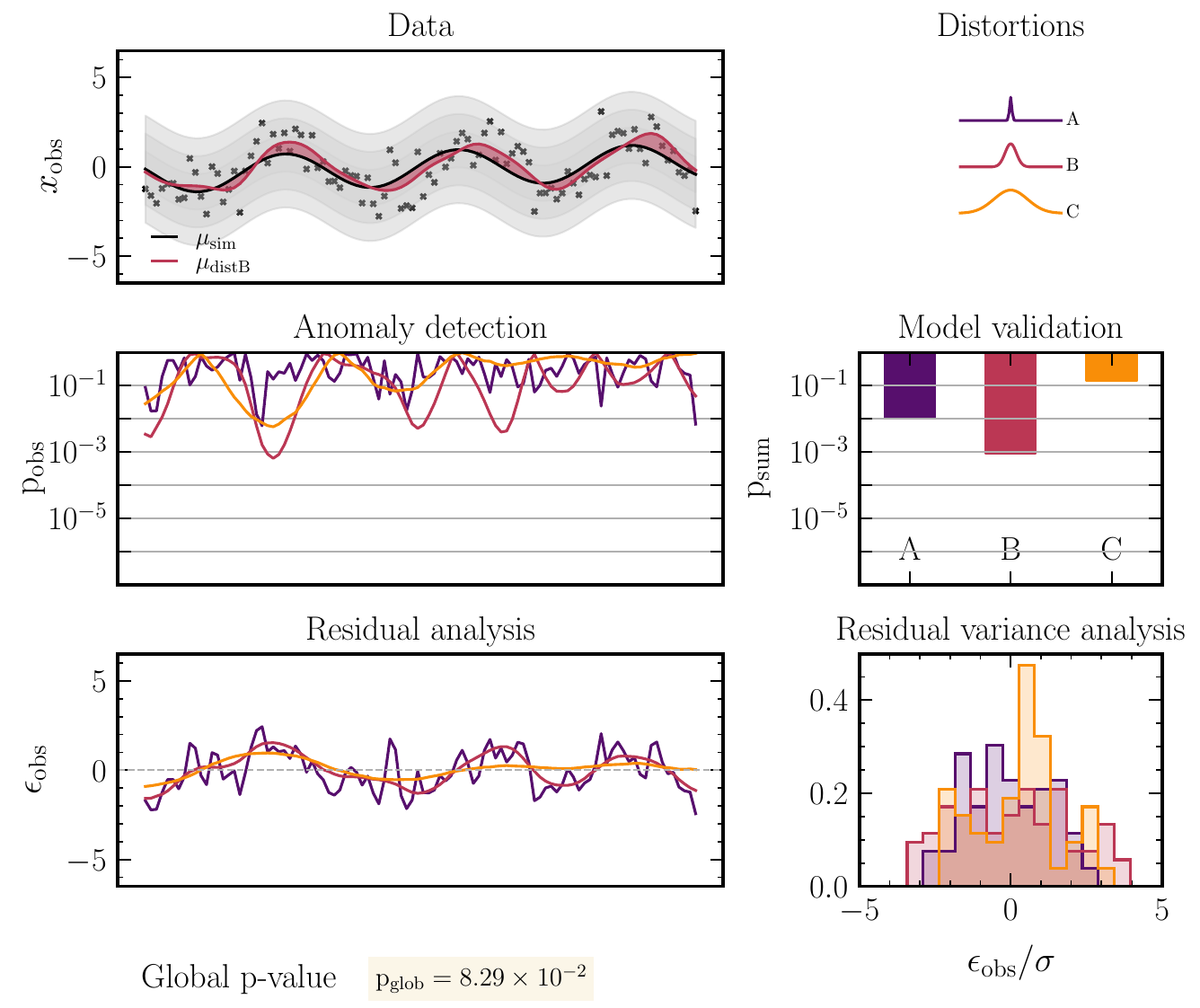}};
    \end{tikzpicture}
    \caption{Same as Figure~\ref{fig:plot1}, but applied to data distorted by many small distortions.}
    \label{fig:plotn}
\end{figure*}

\noindent In this section we show the results of our unified framework for an instructive example: a white noise time series that can be thought of as a toy example for the analysis of \eg\ gravitational waves or light curves. The data is defined across 100 evenly spaced bins, on a grid $y = (y_1, \ldots, y_{100})$ from $y_1 = -10$ to $y_{100} = 10$. Our baseline model consists of unit variance, uncorrelated Gaussian noise in each bin, along with a deterministic signal component that is defined by a sinusoidal function of the grid $y$. Specifically, the sinusoidal function has parameters $\Theta$ that define an adjustable phase, a linear trend, and an offset. Together, these define a mean $\mu_j(\Theta) = \sin(y_j + 0.5\,\Theta_0) +  0.1\,\Theta_1\,y_j + 0.5\,\Theta_2$, where $j$ is the grid index, with $\Theta$ sampled from $p(\Theta) = \mathcal{U}(-1, 1)^3$. The baseline model is thus defined by $p_\text{sim}(\bm{x}) = \int \dd\Theta\ \mathcal{N}(\bm{x}; \bm{\mu}(\Theta), \Sigma = \mathbb{1})\ p(\Theta)$.

We consider a set of three different stochastic additive distortions with three different correlation scales --- dubbed distortions A (spanning 5 bins), B (spanning 21 bins), and C (spanning 61 bins) respectively. We model these correlated distortions with convolutions with kernel sizes corresponding to their correlation scales. By convention, we fix the maximum of the kernels to one. Their position is sampled across the whole data length and their amplitude is sampled from $\epsilon \sim \mathcal{U}(-b, b)$, as in Equation~\ref{eq:additive1}. In Section~\ref{subsec:adaptive} we show how to adaptively learn the strength of the distortions parameter $b$ during training. 

During training, the data is affected by a single distortion at a time and we train in parallel a different network for each different correlation scale. This is to better showcase their differences at inference time, but in principle a single network would suffice for the framework. Here, we adopt the SNR strategy, introduced in Section~\ref{subsec:training}, that allows one to directly estimate the $\mathrm{SNR}$ (linked to the test statistic as in Equation~\ref{eq:t_snr}) by learning the matched filter estimates $\boldsymbol{\epsilon}_\phi(\bm x)$ and their variances $\boldsymbol{\sigma}_\phi^2$ (as defined in Equations~\ref{eq:nnSNR} and~\ref{eq:nnSNRsigma}). A comparison of the SNR and BCE training strategies with the analytical expectation for this example and further details are given in Appendix~\ref{app:comparison}.

In Figures~\ref{fig:plot1} and~\ref{fig:plotn}  we show a summary of our results for a single distortion of type B and for multiple distortions of type B respectively. In the upper left panel of both figures, we show the data points $\bm{x}_\mathrm{obs}$, the baseline signal $\bm{\mu}_\mathrm{sim}$, and the signal distorted by an additive stochastic distortion of type B, $\bm{\mu}_\mathrm{distB}$. The gray shaded areas show the 1-, 2-, and 3-sigma Gaussian noise regions. The rest of the panels highlight the range of results obtained using our framework, and follow the same logic as represented in the summary, Figure~\ref{fig:summary}. These can be described in more detail as follows:

\vspace{5pt}
\noindent \emph{Anomaly detection (central left panel):} This panel shows the anomaly detection significance of localized distortions for the three correlation-scales, shown in the legend (upper right panel). In Figure~\ref{fig:plot1}, we see that there is a clear minimum $\mathrm{p}$-value for the correct correlation scale of the individual distortion. Indeed, we see that the most powerful individual test corresponds to the correct correlation scale at the right point in the data, aligning with matched filtering expectations. In Figure~\ref{fig:plotn}, where there are multiple distortions, we see that we are able to capture their relative significance, and again identify the correct correlation scale.

\vspace{5pt}
\noindent \emph{Residual analysis (bottom left panel):} The residual analysis shows where the distortion is located in data space and how large it is in the same units as the input data. This is enforced by the convention that the maximum element of the kernel of the convolution for the correlated distortion is one. This is shown for the three correlation-scales. It is most useful if there is a localized excess. The behaviour as a residual is perhaps most obvious for the smallest correlation scale, where we see that it tracks upwards and downwards noise fluctuations on top of the signal.

\vspace{5pt}
\noindent \emph{Model validation (central right panel):} The aggregated $\mathrm{p}$-value is useful to check if the model gives a good overall description of the data. It is especially useful when there is no clear single distortion, but many across the data, contributing to the signal, as in Figure~\ref{fig:plotn}. We can also use this to identify the most likely subset of distortions in our data (\eg here the `B' distortion is correctly singled out in Figures~\ref{fig:plot1} and~\ref{fig:plotn} by the analysis).

\vspace{5pt}
\noindent \emph{Residual variance analysis (bottom left panel):} The residual variance analysis helps to visualizes the model validation. In the presence of Gaussian noise it should be standard normally distributed, but the further it is from normality (under the assumptions in Section~\ref{sec:method}), the greater the indication for misspecification.

\vspace{5pt}
\noindent \emph{Global $\mathrm{p}$-value (bottom text):} This is the overall $\mathrm{p}$-value that accounts for the number of tests performed (see Section~\ref{subsec:global}). The $\mathrm{p}$-values from the localized and aggregated tests in the second rows are individual $\mathrm{p}$-values for specific tests. Here, we choose to treat the aggregated and localized tests on equal footing by combining them into a single global $\mathrm{p}$-value, which accounts for all 303 tests under consideration (3 aggregated tests and 300 localized tests). Of course, one could opt to derive global $\mathrm{p}$--values separately for the localized and aggregated tests if desired. This choice can be adapted to the specific testing framework or interpretational needs.

\subsection{Self-calibrating distortions algorithm} \label{subsec:adaptive}

\begin{figure*}
    \centering
    \includegraphics[width=\linewidth,clip]{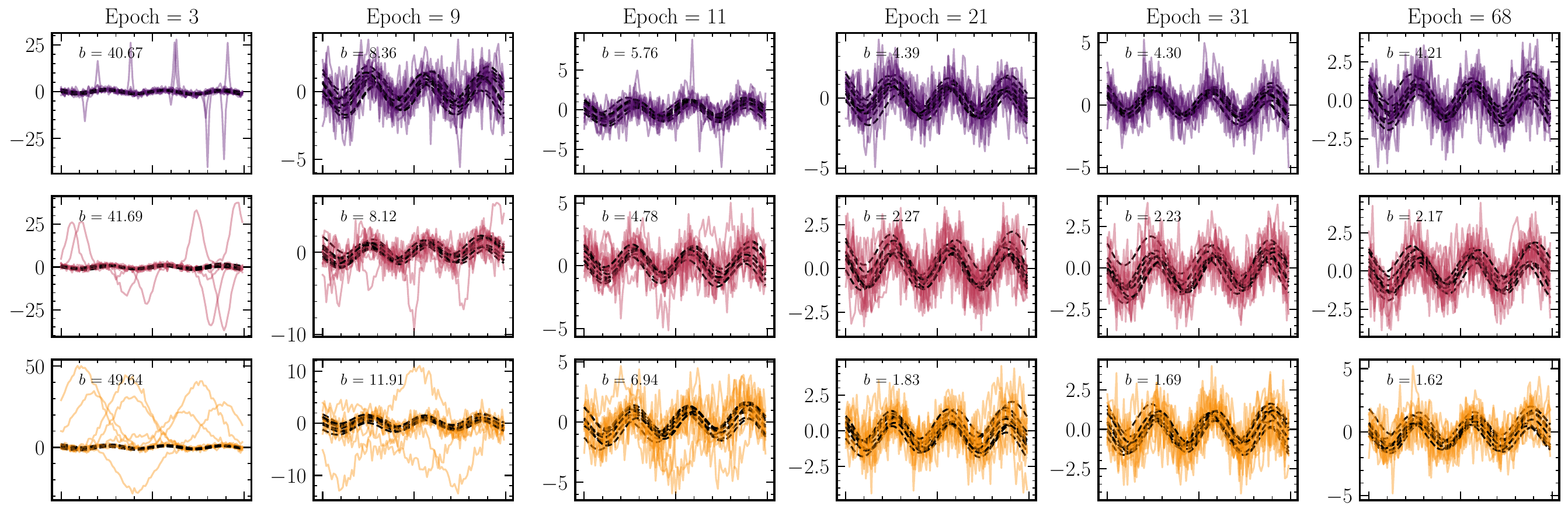}
    \caption{Illustration of the adaptive training of distortion amplitudes in our framework. The figure shows how the generated distortions (color-coded to match the legend in Figure~\ref{fig:plot1}) dynamically adjust to envelop the deterministic part of the baseline model (black dashed lines) during training. The baseline model noise is not shown in the plot for clarity purposes. By adaptively tuning the distortion amplitude parameter $b$ based on the learned variance $\sigma^2$ and a desired maximum signal-to-noise ratio $\mathrm{SNR}_\mathrm{max}$, the distortions remain plausible.}
    \label{fig:adaptive}
\end{figure*}

\noindent An important consideration in setting up this framework is determining the variance of the distortion amplitude, as defined by the parameter $b$ (see Equation~\ref{eq:additive1}). Ideally, with sufficiently flexible networks and infinite training data, the test results would be largely invariant to the specific variations of $b$, as long as it is not too small. In practice, however, choosing distortions that are only slightly larger than the natural statistical baseline model variations results in more efficient training. 

In simple cases one can make an educated guess about $b$ or even calculate it analytically, but it is a straightforward extension of the SNR training strategy (Appendix~\ref{app:SNR}) to make it adaptive so as to converge only to plausible distortion amplitudes. In other words, it converges to distortions that are significant enough to be detectable, but not so significant that they are clearly ruled out. One way to make this quantitative is to define a maximum SNR for a given distortion, $\mathrm{SNR}_\mathrm{max}$. This is realized by the maximum value of $\epsilon$, \ie\ $b$. Writing this in terms of the variance $\sigma$, we see that:
\begin{equation}
    b = \text{SNR}_\text{max} \sigma = \frac{\text{SNR}_\text{max}}
    {\sqrt{(\bm n^{(i)})^T \Sigma^{-1} \bm n^{(i)}}}
    \;.
\end{equation}
Since $\sigma^2$ is a learned parameter of the model in the SNR training strategy, we can anchor $b$ to its learned value on the fly by choosing a desired $\text{SNR}_\text{max}$ (for our example we set $\text{SNR}_\text{max}=5$), and generate training data with the more suitable distortion amplitude as the network is learning.

Figure~\ref{fig:adaptive} shows this in practice, highlighting how the distortions correctly end up enveloping the model.~\footnote{For simplicity, since in this example the variance $\sigma^2$ is pretty much constant across distortions of the same correlation scale, we estimate a single $b$ for all localized distortions of the same correlation scale, anchoring it to the mean value of $\sigma^2$.} This adaptive approach ensures that the alternative hypotheses explored during high-volume hypothesis testing are both challenging and realistic, enhancing the sensitivity of our method to subtle model discrepancies.

\section{Application to gravitational waves} \label{sec:gw}

\noindent In this section, we move beyond the toy problem that we have investigated in detail above and use our framework to analyse real data. In particular, we take the example of gravitational waves (GWs), as detected by the LIGO-Virgo-Kagra collaboration. Specifically, we will focus on the first detection by the LIGO detectors~\cite{LIGOScientific:2016aoc,LIGOScientific:2016vlm,LIGOScientific:2016vbw}, GW150914, and illustrate how to perform an additional post-analysis quality check with our framework. Beyond this, our framework could be used to test other various aspects of GW data analysis pipelines. For example, it could be used to test for waveform systematics after carrying out Bayesian inference (see \eg~\cite{LIGOScientific:2016ebw,Gamba:2020wgg,Lam:2023oga} for discussions on systematics in the LIGO context), compare and contrast different models for detector noise in the presence/absence of astrophysical signals~\cite{LIGOScientific:2016emj,Legin:2024gid}, or identify and flag detector artifacts or glitches (see \eg~\cite{Coughlin:2019ref}) that may result in a mismodelling or biasing of the parameter estimation. However, we postpone a full follow up of this application to a future work.

\subsection{Bayesian inference step} \label{subsec:gw_inference}

\noindent The starting point of our analysis is the circumstance where we have fitted some model to data, and are then looking to evaluate whether there is any evidence for mismodelling. In the GW context, we carry out a likelihood-based Bayesian inference analysis of the first LIGO event, GW150914~\cite{LIGOScientific:2016aoc,LIGOScientific:2016vlm,LIGOScientific:2016vbw}. From a technical point of view, this involves \emph{(a)} defining an analysis window and accessing the data from the GWOSC\footnote{Gravitational Wave Open Science Center}~\cite{LIGOScientific:2019lzm}, \emph{(b)} estimating the power-spectral density $S_n(f)$ (PSD) around the event, \emph{(c)} choosing a signal model to fit to the data using an appropriate likelihood and sampler. 

In this work, we use the \texttt{jimgw} library~\cite{Wong:2023lgb}, built on top of the \texttt{ripple} waveform package~\cite{Edwards:2023sak} to analyse GW150914, with the priors and detector setups exactly as described in Ref.~\cite{Wong:2023lgb}. As far as \emph{(a)} is concerned, we analyse the data from both the Hanford (H1) and Livingston (L1) detectors, with a $4\, \mathrm{s}$ detection window centred on a GPS trigger time of $1126259462.4$, sampled at 4096 Hz. For \emph{(b)}, we estimate the PSD using a data segment of length $16 \,\mathrm{s}$ starting $32\,\mathrm{s}$ before the beginning of the detection window. Finally, for \emph{(c)}, we use the \texttt{IMRPhenomD} waveform model~\cite{Husa:2015iqa,Khan:2015jqa}, as implemented in the \texttt{ripple} codebase~\cite{Edwards:2023sak}. The result of this is a set of posterior samples $\Theta \sim p(\Theta | h_{\mathrm{H}_1}, h_{\mathrm{L}_1})$ over the gravitational wave model parameters $\Theta$~\footnote{Here, $\Theta$ consists of all the standard intrinsic (as relevant to the \texttt{IMRPhenomD} waveform model~\cite{Husa:2015iqa,Khan:2015jqa}) and extrinsic parameters describing the properties and location of the GW source.} given the detector data $h_{\mathrm{H}_1}, h_{\mathrm{L}_1}$ for Hanford and Livingston, respectively. These (noise-free) posterior-predictive samples are visualised (after the data processing steps described below) in the top panel of Figure~\ref{fig:gw150914} alongside the real (processed) data for the Hanford detector.

\subsection{Data processing setup} \label{subsec:gw_process}

\noindent For this application, we use our model misspecification framework to check for deviations from the posterior-predictive distribution obtained in the inference step. We do this for whitened time domain data $d_w(t)$ in the Hanford detector as a concrete example. 

To generate simulated data in the time domain under our null hypothesis model $H_0$, we use the following procedure. First, we take a posterior sample $\Theta \sim p(\Theta | h_{\mathrm{H}_1}, h_{\mathrm{L}_1})$ and generate the frequency domain signal $\tilde{h}_{H_1}(f; \Theta)$ using the \texttt{IMRPhenomD} waveform model. In addition, we generate noise in the frequency domain $\tilde{n}(f)$ by sampling from the PSD estimated during the inference step. Then, before taking the inverse Fast Fourier Transform (FFT), we normalize the frequency domain strain by the square-root of the PSD and filter the data using a bandpass filter between $20 \, \mathrm{Hz}$ and $1024 \, \mathrm{Hz}$~\footnote{Note that this is the same frequency range used during the inference step~\cite{Wong:2023lgb}.}, with additional notches of a width $0.1 \, \mathrm{Hz}$ removed at $60$, $120$, and $180 \, \mathrm{Hz}$. For visualization purposes, we also downsample the resulting time domain data by a factor 8 and consider a time window of 0.2 s around the trigger time. To summarise in the language of Section~\ref{sec:method}, our baseline model in this context is given by $p_\text{sim}(d_w(t)) = \int \dd\Theta\ p(d_w(t)| \Theta, S_n) p(\Theta | h_{\mathrm{H}_1}, h_{\mathrm{L}_1})$. It is worth stressing that using posterior samples here is somewhat for illustrative purposes, since it is expected that this will slightly overestimate distortions that are degenerate with the effect of model parameter changes. In general, parameters should be drawn from the full prior or truncated versions thereof~\cite{AnauMontel:2023stj,miller_truncated_2021}.

To analyse the real data from the Hanford detector for GW150914, there is one additional step that we must carry out. In particular, before taking the FFT of the raw time domain data to move to the frequency domain where the whitening and filtering is carried out, we must apply a window function. Here, we apply the same Tukey window function as in the original Bayesian inference analysis (see the \texttt{jimgw} code~\cite{Wong:2023lgb}). 

\subsection{Results} \label{subsec:gw_results}

\begin{figure}
    \centering
    \includegraphics[width=\linewidth,clip]{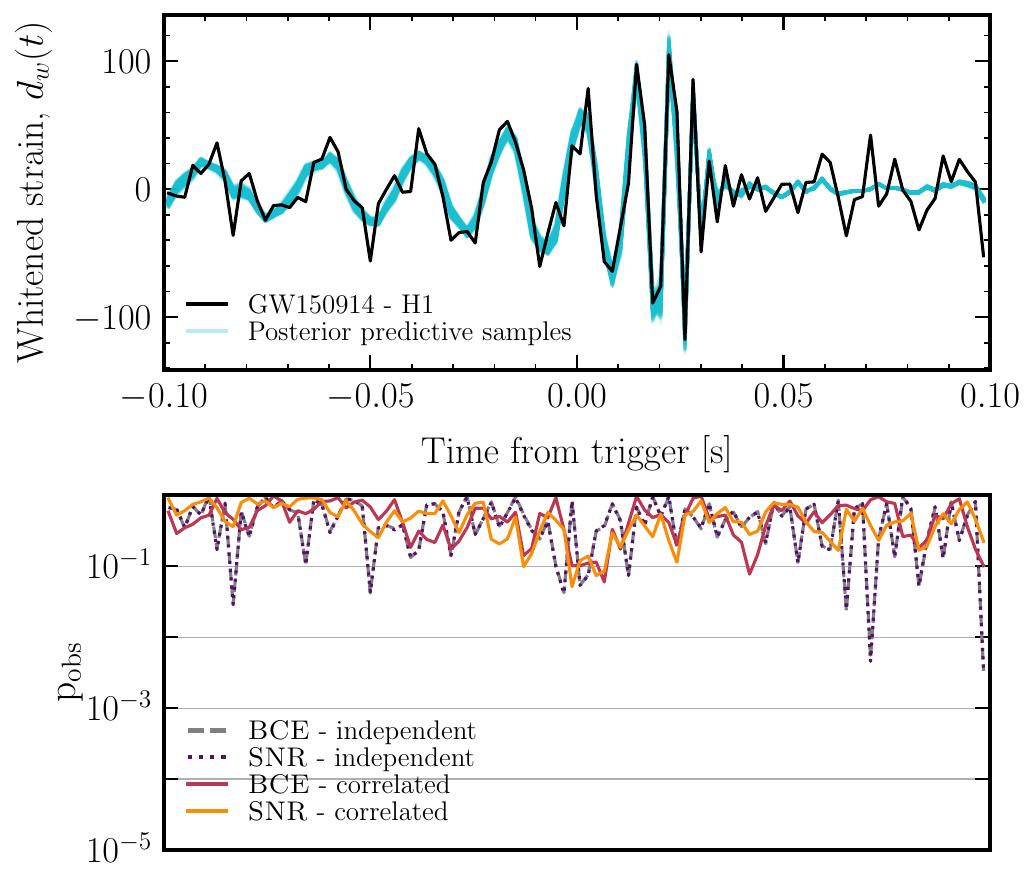}
    \caption{\textbf{Top panel:} GW150914 data for the Hanford detector and posterior-predictive distribution samples from the Bayesian inference step as described in Section~\ref{subsec:gw_inference} and processed as described in Section~\ref{subsec:gw_process}.
    \textbf{Bottom panel:} Results of our framework for GW150914 from the Hanford detector. We test for an independent bin-wise distortion and a correlated one, using both our training strategies, dubbed BCE and SNR respectively (Section~\ref{subsec:training}). 
    As expected, no significant anomaly is present in the modelling of GW150914, with global $\mathrm{p}$-values for all the types of analyses of around a few tenths.}
    \label{fig:gw150914}
\end{figure}

\noindent For the purpose of this example, our baseline model is the one described in the previous section. As deviations from the base simulator, we consider two types of stochastic additive distortions: an independent, bin-wise distortion for each processed time step (totally 102 bins), and a correlated distortion spanning eleven processed time steps generated through a convolution in the same way as the one for the example in Section~\ref{sec:example}.

For both classes of distortions, we tested both our training strategies, BCE and SNR (Section~\ref{subsec:training}). The results are presented in Figure~\ref{fig:gw150914}. As expected, no significant anomaly is present in the modeling of GW150914, with global $\mathrm{p}$-values for all the types of analyses of around a few tenths. We do not report a plot for the aggregated test statistic for model validation since it also shows similarly a very small significance. 

We have also tested the results against different processing steps of our data. In particular, we also tried using unwhitened data, with a 40-400 Hz bandpass filter, or downsampling by a factor of 2 instead of 8. In the former case, this introduces the non-trivial feature that the noise in the time domain is correlated. Nonetheless, in all of these additional cases the results were consistent with those shown in Figure~\ref{fig:gw150914}.

Note that the example application that we have demonstrated here is only one option as far as GW data analysis is concerned, although the application to real data is an important step. As discussed at the start of this section, it would be interesting to take this further and study either a broader class of events, or look to calibrate detector performance and waveform models.

\section{Discussion} \label{sec:discussion}

\noindent This work focuses on detecting a wide variety of model discrepancies through high-volume hypothesis testing. However, we do not address how to adjust inference algorithms to accommodate these discrepancies if they are identified, nor do we examine the robustness of estimators in the presence of systematic errors.
We acknowledge several recent efforts aimed at the first issue --- specifically, adapting inference algorithms trained on one simulator for application to another.
These approaches involve selecting subsets of observables that remain relatively consistent across different simulators \cite{echeverri2023cosmologygalaxyastrid, de_Santi_2023}, employing domain adaptation techniques during the training phase \cite{swierc2024domainadaptiveneuralposteriorestimation, Lee_2024}, or correcting minor inaccuracies in the simulations through calibration methods \cite{jia2024cosmologicalanalysiscalibratedneural}. Regarding the second issue, the robustness of SBI techniques, such as neural posterior estimation and neural ratio estimation, to distributional shifts was recently investigated by Ref.~\cite{filipp2024robustnessneuralratioposterior}.

With this context in mind, the main advantage of high-volume hypothesis testing for many test statistics $t_i$, localized in the space of distortions, is that any mismodelling can be systematically identified and visualized, as \eg\ in Figure~\ref{fig:plot1}. An aggregated test statistic on the other hand can only reveal that some part of the data is mismodeled, but it cannot tell us exactly how.
By directly targeting localized test statistics $t_i$, our method opens up a plethora of hypothesis tests for the same trained network, since the way we combine them into aggregated tests is entirely flexible. In Section~\ref{sec:method}, we highlighted the summed test statistic given by $t_\mathrm{sum} = \sum_i t_i$. However, in principle many other options are open. For example, we could define aggregated tests for specific subsets of the distortions (\eg, focusing on distortions at a particular correlation scale, or on the left half of the data space etc.), or construct complex statistics like a ``double-excess" test statistic for the probability of two large excesses anywhere in the data. Such a construction, which would be analytically challenging, is now trivial to implement and to compute the significance for, by summing only the largest and second-largest localized test values.

On the other hand, this flexibility does come with computational costs, as the dimensionality of the network output scales with the number of alternative hypotheses (see \eg~Equations~\ref{eq:nnBCE} and \ref{eq:nnSNR}). Hence, when looking for a wider variety of distortions, it could be preferable to directly target global test statistics, or test statistics at the level of informative lower dimensional summaries of the data.

One of the main benefits of our framework is that the choice of sampling distributions for the alternative hypotheses (or equivalently the form of the model augmentation $\bm n$) can be tailored to the questions of interest. Whilst, in principle, any choice will lead to a concrete set of test statistics, there are likely to be trade-offs between complexity, specificity, and functional form of the contrastive distribution. Consequently, the implications of these choices on the statistical power of the test to detect significant features will vary on a case by case basis. Regardless, the adaptive algorithm described in Section~\ref{subsec:adaptive} can be employed to set the variations of the strength of the distortions, so as they are plausible. Furthermore, we note that this model augmentation need not occur in data space directly. For example, in gravitational wave analyses, the augmentation could be performed at the signal level in frequency space, offering additional flexibility in defining meaningful alternative hypotheses.

As a last point of discussion, we expand upon our findings regarding the main upsides and downsides of the two training strategies (Section~\ref{subsec:training}). We note that the followings are not exhaustive results, but intuitions from our experiments that can serve as useful starting point for further investigation. By construction, the SNR training strategy provides clear and interpretable results, allowing one to visually inspect the amplitude of deviations in specific distortion directions through $\epsilon$ (see \eg\ the bottom left panel of Figure~{\ref{fig:plot1}}). Furthermore, it allows one to adaptively learn during training the most appropriate amplitude for the distortions of interest (Section~\ref{subsec:adaptive} and Figure~\ref{fig:adaptive}). On the other hand, the BCE training strategy does not depend on any prior assumption. As such, we expect it to better perform in general scenarios. Indeed, we have found that it seems to give a slightly better performance when describing excesses and localizing distortions, as shown, \eg, in Figure~\ref{fig:comparison}.

\section{Conclusion} \label{sec:conclusion}

\noindent Model misspecification analysis strategies are integral to advancing our understanding of physical phenomena. The framework presented here is designed to carry this out in a simulation-based inference context. By leveraging classical concepts, it provides a flexible and comprehensive approach to simultaneously perform many hypothesis tests and quantify their statistical significance (Section~\ref{sec:method}). In Section~\ref{sec:example}, we demonstrated the application of the framework to a toy example, before applying it to real data in the gravitational wave context in Section~\ref{sec:gw}. The main conclusions of this work are then as follows:

\vspace{5pt}
\noindent \emph{Detection of model misspecification:} Our framework uses high-volume hypothesis testing to detect model misspecification. This framing allows us to unify the ideas of anomaly detection (localized test statistics) and model validation (aggregated tests), as described in Section~\ref{sec:method}. In addition, due to its simulation-based nature, it also is extremely flexible as far as the classes and types of distortion that can be searched for.

\vspace{5pt}
\noindent \emph{Efficient:} As described in Section~\ref{sec:method}, one way to think about our framework is as a collection of hypothesis tests comparing the various distorted data models to a base simulator. Via the training strategies described in Section~\ref{subsec:training}, we can actually test (and Monte Carlo sample) all of these alternative hypotheses simultaneously. This makes the pipeline very efficient when looking to test for broad classes of mismodelling, while still maintaining the ability to carry out individual, targeted tests. Furthermore, for the SNR training strategy, we have developed an adaptive algorithm that can be used to calibrate the scale of distortions searched for in the data (Section~\ref{subsec:adaptive}). 

\vspace{5pt}
\noindent \emph{Principled:} Although the method is simulation-based, it is firmly rooted in classical statistical principles. Indeed, we discussed at length the connections to classical hypothesis testing in Section~\ref{sec:method}. For example, we showed how our framework reduces to the classical concepts of matched filtering and $\chi^2$ goodness-of-fit tests under certain conditions. This adds an additional level of interpretability to the results derived in this work.

\vspace{5pt}
\noindent To conclude, we argued at the start of this work that there was a crucial need for model misspecification frameworks in the context of SBI. The presented work may serve as a step towards that direction. This will allow us to move beyond the parameter estimation regime, and encourage more end-to-end, simulation-based analysis pipelines for real-world data settings across astrophysics, particle physics, and cosmology.

\section*{Acknowledgements}

\noindent The work of NAM, CW, and JA was supported by a project that has received funding from the European Research Council (ERC) under the European Union’s Horizon 2020 research and innovation program (Grant agreement No.~864035 – UnDark). JA is supported by a fellowship from the Kavli Foundation.

\bibliography{references}

\newpage
\clearpage

\appendix
\makeatletter
\renewcommand{\thesubsection}{\Alph{section}.\arabic{subsection}}
\makeatother

\noindent \textbf{APPENDICES}

\vspace{10pt}
\noindent Appendix~\ref{app:BCE} describes a general training strategy for our model misspecification testing framework. An alternative training strategy, that stems from the connection of the framework to classical testing (Appendix~\ref{app:math}), is presented in Appendix~\ref{app:SNR}. Furthermore, a comparison of the two strategies to each other and the analytical counterpart is presented in Appendix~\ref{app:comparison}.

\section{Classifier-based training strategy}\label{app:BCE}

\noindent As discussed in Section~\ref{sec:method}, we want to approximate an ensemble of $N_\mathrm{alt}$ test statistics $t_i$ (Equation~\ref{eq:ti}) via neural networks. As originally proposed in Ref.~\cite{cranmer2016approximating}, discriminative classifiers can be used to approximate the generalized likelihood ratio statistic when only a generative model for the data is available.
The classifiers $f_{i, \phi}$ can be optimized through gradient descent using the standard binary cross-entropy loss \cite{mao2023crossentropylossfunctionstheoretical} as the optimization objective,
\begin{multline}
    \mathcal{L}^{(i)}_\mathrm{BCE}\left[f_{i, \phi}(\bm x)\right]  
    =
    \mathbb{E}_{\bm{x} \sim p_\mathrm{sim}(\bm{x})} \left[-\ln{\sigma{(f_{i, \phi}(\bm x))}}\right]+\\
    \mathbb{E}_{\bm{x} \sim p_\mathrm{dist}(\bm{x} | i)} \left[-\ln{\sigma{(1-f_{i, \phi}(\bm x))}}\right]
    \;,
\end{multline} 
although other classes of loss functions could also be employed~\cite[\eg][]{jeffrey_evidence_2024}.

After optimisation, each classifier estimates the likelihood ratio
\begin{equation} \label{eq:classifier}
    f_{i, \phi}(\bm x) \approx \ln\frac{p_\mathrm{dist}(\bm{x} | i)}{p_\mathrm{sim}(\bm{x})}\;,
\end{equation}
and we can define the localized test statistics in terms of the its output via
\begin{equation} \label{eq:t_class}
    \hat{t}_i(\bm{x}) = 2 f_{i, \phi}(\bm x) \simeq - 2 \ln\frac{p_\mathrm{sim}(\bm{x})}{p_\mathrm{dist}(\bm{x} | i)}\,.
\end{equation}

In the interest of efficiency, rather than training individual classifiers for each possible localized distortion $i$, we typically train single multi-output networks of the form
\begin{equation} \label{eq:nnBCE}
    \bm{f}_\phi(\bm x): \mathcal{D} \to \mathbb{R}^{N_\text{alt}}
    \;,
\end{equation}
where $\mathcal{D}$ refers to the data space of $\bm x$. The total loss is given as a sum over the individual losses,
\begin{equation}
    \mathcal{L}_\mathrm{BCE}\left[\bm{f}_{\phi}(\bm x)\right]=
    \sum^{N_\text{alt}}_{i=1} \mathcal{L}^{(i)}_\mathrm{BCE}[f_{i, \phi}(\bm x)] \;,
\end{equation}
where the sum runs over all hypotheses $H_i$.

\section{Connection to classical testing frameworks} \label{app:math}

\noindent In this appendix, we derive the connection between the proposed distortion-driven model misspecification testing, presented in Section~\ref{sec:method}, and classical testing frameworks. Throughout this derivation, a number of assumptions (highlighted in bold) will be made in order to align our general framework with more specific traditional testing methods.

To begin, we note that traditional tests often use \emph{profiled likelihoods}, whereas our method considers \emph{likelihoods marginalized over model parameters}. Thus, in order to establish the connection between our method and classical testing frameworks, we first compute the relationship between marginal and profile likelihoods for $H_0$ and $H_i$. 

\vskip 10pt
\noindent \textbf{Marginal null hypothesis.} Given a model likelihood $p(\bm x |\Theta)$, our base hypothesis is defined by \emph{marginalizing} over its parameters $\Theta$
\begin{equation}
    p(\bm x | H_0) = 
    \int d\Theta\; p(\Theta) p(\bm x | \Theta) 
\end{equation}
where $p(\Theta)$ is the prior distribution over the model parameters. 
In contrast, the \emph{profile} likelihood $p(\bm x | \Theta^\ast_{\bm x})$, is the value of the likelihood at its maximum-likelihood estimator (MLE)
\begin{equation} \label{eq:mle}
    \Theta^\ast_{\bm x} = \arg\max_{\Theta} p(\bm x | \Theta)\;.
\end{equation}

To connect marginal and profile likelihoods, we assume the \textbf{large sample limit}, where it is possible to approximate the likelihood function as Gaussian in $\Theta$,\footnote{Note that this does not require that $p(\bm x | \Theta)$ is Gaussian in $\bm x$.} centered around the MLE $\Theta^\ast_{\bm x}$ and with covariance $\Sigma_{\Theta^\ast_{\bm x}}$
\begin{equation}\label{eq:large_sample_limit}
    p(\bm x | \Theta) \propto \mathcal{N}(\Theta; \Theta^\ast_{\bm x} , \Sigma_{\Theta^\ast_{\bm x}}) \;.
\end{equation}

Under the above assumption, we can connect marginal and profile likelihoods as follows
\begin{equation} \label{eq:pxH0}
\begin{split}
    p(\bm x &| H_0)
    =  p(\bm x | \Theta^\ast_{\bm x})  \int d\Theta\; p(\Theta) \frac{p(\bm x | \Theta)}{p(\bm x | \Theta^\ast_{\bm x})}\\
    &\stackrel{\text {\ref{eq:large_sample_limit}}}{\simeq}  p(\bm x | \Theta^\ast_{\bm x}) \times \E_{\Theta \sim \mathcal{N}(\Theta^\ast_{\bm x}, \Sigma_{\Theta^\ast_{\bm x}})}p(\Theta)
    \sqrt{(2\pi)^d \det \Sigma_{\Theta^\ast_{\bm x}}} \;.
\end{split}
\end{equation}

\vskip 10pt
\noindent \textbf{Marginal alternative hypothesis.} Let us now consider, as alternative hypotheses $H_i$, the case where the distortions are in the form of simple \textbf{stochastic additive non-Gaussian distortions} in specific noise directions $\bm{n}^{(i)}$,
\begin{equation} \label{eq:additive}
    H_i: 
    \tilde{\bm{x}} = \bm{x} + \epsilon \cdot \bm{n}^{(i)} \quad \text{with} \quad \epsilon \sim \mathcal{U}(-b, b)\;,
\end{equation}
where $\bm{x} \sim p(\bm{x}| H_0)$.
Note that, although $\epsilon$ is a random variable, the noise directions $\bm{n}^{(i)}$ are considered to be \textit{fixed}. The bounds $b$ for the prior of $\epsilon$ are chosen large enough so that $H_i$ is significantly different from $H_0$, while specific choices will be discussed below in Section~\ref{subsec:boundaries}.

In this case, the likelihood for the $H_i$ hypothesis can be expressed as a convolution of the likelihood under $H_0$ via,
\begin{equation}
\begin{split}
    p(\bm x &| H_i) = \int d\epsilon \; 
    p(\epsilon) p(\bm x - \epsilon \bm n^{(i)} | H_0) \\
    &= \int d\epsilon \; 
    p(\epsilon) \int d\Theta\; p(\Theta) p(\bm x - \epsilon \bm n^{(i)} | \Theta)\;.
\end{split}
\end{equation}
Following the same steps as in the previous section to derive Equation~\ref{eq:pxH0} for $p(\bm x | \Theta)$, it is possible to connect the marginal and the profile likelihood of $p(\bm x - \epsilon \bm n^{(i)} | \Theta)$ assuming the large sample limit. Here, to facilitate our ultimate goal, \ie\ to compute the ratio between $p(\bm x| H_0)$ and $p(\bm x| H_i)$ for the test statistic, we consider the additional assumption that the \textbf{MLE of $p(\bm x - \epsilon \bm n^{(i)} | \Theta)$ is not significantly correlated with the distortions $\epsilon \bm n^{(i)}$}. In other words, we assume that the best-fit value for $\Theta$ is not significantly affected by a single extra noise degree of freedom
\begin{equation}\label{eq:not_affected}
    \Theta^\ast_{\bm x} \simeq \Theta^\ast_{\bm x - \epsilon \bm n^{(i)}} \;.
\end{equation}
With the above assumption, we can derive,
\begin{equation}
\begin{split}
    p(\bm x | H_i) &= \int d\epsilon \; 
    p(\epsilon) \int d\Theta\; p(\Theta) p(\bm x - \epsilon \bm n^{(i)} | \Theta) \\
    &\stackrel{\text {\ref{eq:not_affected}}}{=} \int d\epsilon \; 
    p(\epsilon) p(\bm x-\epsilon \bm n^{(i)} | \Theta^\ast_{\bm x}) \\ & \qquad \int d\Theta\; p(\Theta) \cfrac{p(\bm x - \epsilon \bm n^{(i)} | \Theta)}{p(\bm x-\epsilon \bm n^{(i)} | \Theta^\ast_{\bm x})} \\
   &\stackrel{\text {\ref{eq:large_sample_limit}, \ref{eq:not_affected}}}{\simeq}  \int d\epsilon\; p(\epsilon)
   p(\bm x-\epsilon \bm n^{(i)} | \Theta^\ast_{\bm x}) \\
   & \qquad \times \E_{\Theta \sim \mathcal{N}(\Theta^\ast_{\bm x}, \Sigma_{\Theta^\ast_{\bm x}})}p(\Theta) \sqrt{(2\pi)^d \det \Sigma_{\Theta^\ast_{\bm x}}} \;.
\end{split}
\end{equation}
We can further simplify the above expression by introducing the MLE for $\epsilon$, $\epsilon^\ast_{\bm x}$, and the second derivative around the curvature, $\sigma_{\epsilon^\ast_{\bm x}}^{-2}$,
\begin{flalign} 
    \epsilon^\ast_{\bm x} &\equiv \arg\max_{\epsilon} p(\bm x - \epsilon \bm n^{(i)} | \Theta^\ast_{\bm x}) \;, \label{eq:epsilon}\\
    \sigma_{\epsilon^\ast_{\bm x}}^{-2} &\equiv \partial^2_\epsilon \ln p(\bm x - \epsilon \bm n^{(i)})|_{\epsilon = \epsilon^\ast_{\bm x}} \label{eq:sigma_epsilon}\;.
\end{flalign}
We then obtain
\begin{equation} \label{eq:pxH1}
\begin{split}
    p(\bm x &| H_i)
    \simeq  {p(\bm x-\epsilon^\ast_{\bm x} \bm n^{(i)} | \Theta^\ast_{\bm x})} \\ & \qquad \times \int d\epsilon\; p(\epsilon)
   \cfrac{p(\bm x-\epsilon \bm n^{(i)} | \Theta^\ast_{\bm x})}{p(\bm x-\epsilon^\ast_{\bm x} \bm n^{(i)} | \Theta^\ast_{\bm x})} \\
   & \qquad \times \E_{\Theta \sim \mathcal{N}(\Theta^\ast_{\bm x}, \Sigma_{\Theta^\ast_{\bm x}})}p(\Theta) \sqrt{(2\pi)^d \det \Sigma_{\Theta^\ast_{\bm x}}} \\
   &\simeq  {p(\bm x-\epsilon^\ast_{\bm x} \bm n^{(i)} | \Theta^\ast_{\bm x})} p(\epsilon^\ast_{\bm x}) \sqrt{2\pi  \sigma_{\epsilon^\ast_{\bm x}}^2} \\
   & \qquad \times \E_{\Theta \sim \mathcal{N}(\Theta^\ast_{\bm x}, \Sigma_{\Theta^\ast_{\bm x}})}p(\Theta) \sqrt{(2\pi)^d \det \Sigma_{\Theta^\ast_{\bm x}}} \;,
\end{split}
\end{equation}
where, in the second step, the integral over $\epsilon$ is approximated by assuming that $p(\bm x-\epsilon \bm n^{(i)} | \Theta^\ast_{\bm x})$ is sharply peaked around $\epsilon^\ast_{\bm x}$, as expected in the large sample limit.

\vskip 10pt
\noindent \textbf{Marginal test statistic.} We have now all the elements to compute the test statistic quantity of interest (Equation~\ref{eq:ti}), 
\begin{equation} \label{eq:ti_step1}
\begin{split}
    t_i(\bm x) &= -2\ln \cfrac{p(\bm x | H_0)}{p(\bm x | H_i)} \\
    &\stackrel{\text{\ref{eq:pxH0},\ref{eq:pxH1}}}{\simeq} -2\ln \cfrac{p(\bm x | \Theta^\ast_{\bm x})}{p(\bm x-\epsilon^\ast_{\bm x} \bm n^{(i)} | \Theta^\ast_{\bm x})p(\epsilon^\ast_{\bm x}) \sqrt{2\pi  \sigma_{\epsilon^\ast_{\bm x}}^2}}\\
\end{split}
\end{equation}
To connect the above quantity to classical analysis frameworks, we consider their common assumption of a \textbf{Gaussian likelihood function}, and define
\begin{flalign}
    p(\bm x | \Theta) &\equiv \mathcal{N}(\bm x; \bm \mu(\Theta), \Sigma) \;, \label{eq:gauss1} \\
    p(\bm x - \epsilon \bm n^{(i)} | \Theta) & \equiv \mathcal{N}(\bm x- \epsilon \bm n^{(i)}; \bm \mu(\Theta), \Sigma) \label{eq:gauss2}
\end{flalign}
where $\bm \mu(\Theta)$ is the model prediction and $\Sigma$ the likelihood covariance matrix. 

Given the Gaussian assumption, we can straightforwardly compute the value of the MLE of $\epsilon$ and its variance
\begin{flalign}
    \epsilon^\ast_{\bm x} &\stackrel{\text{\ref{eq:epsilon},\ref{eq:gauss2}}}{=} 
    \frac
    {\Delta \bm x^T \Sigma^{-1} \bm n^{(i)}}
    {(\bm n^{(i)})^T \Sigma^{-1} \bm n^{(i)}}  \label{eq:epsilon_gauss} \\
     \sigma_{\epsilon^\ast_{\bm x}}^{-2} &\stackrel{\text{\ref{eq:sigma_epsilon},\ref{eq:gauss2}}}{=}{(\bm n^{(i)})^T \Sigma^{-1} \bm n^{(i)}}  \label{eq:sigma_gauss}
\end{flalign}
where we have defined the residual between the data and the maximum-likelihood model prediction, 
\begin{equation} \label{eq:deltax}
    \Delta \bm x \equiv \bm x - \bm \mu(\Theta^\ast_{\bm x})\;.
\end{equation}
Finally, in the case of a Gaussian likelihood with non-Gaussian additive distortions, assuming the large sample limit, and that the model parameter MLE is not significantly affected by the distortions, we obtain that the marginal likelihood ratio can be written as
\begin{equation} \label{eq:tSNR}
        t_i(\bm x) = -2\ln \frac{p(\bm x | H_0)}{p(\bm x | H_i)} \stackrel{\footnotemark}{\simeq} \text{SNR}_i^2(\bm x) + C
\end{equation}
\addtocounter{footnote}{-1}
\footnotetext{
    Expanding the computations
    \begin{equation*}
    \begin{split}
        t_i &\stackrel{\text{\ref{eq:ti_step1},\ref{eq:gauss1},\ref{eq:gauss2},\ref{eq:deltax}}}{\simeq} - (\Delta\bm x - \epsilon^\ast_{\bm x}\bm n^{(i)})^T \Sigma^{-1}(\Delta\bm x - \epsilon^\ast_{\bm x}\bm n^{(i)}) \\ 
        & \quad + \Delta\bm x^T \Sigma^{-1}\Delta \bm x + 2\ln p(\epsilon^\ast_{\bm x})+\ln(2\pi\sigma^2_{\epsilon^\ast_{\bm x}})\\
        & \stackrel{\text{\ref{eq:epsilon_gauss}}}{=}  \left(\frac{\Delta \bm x^T \Sigma^{-1} \bm n^{(i)}}{\sqrt{(\bm n^{(i)})^T \Sigma^{-1} \bm n^{(i)}}}\right)^2 + 2\ln p(\epsilon^\ast_{\bm x})+\ln(2\pi\sigma^2_{\epsilon^\ast_{\bm x}})\\
        &\stackrel{\text{\ref{eq:epsilon_gauss},\ref{eq:sigma_gauss}}}{=}  \left(\frac{\epsilon^\ast_{\bm x}}{\sigma_{\epsilon^\ast_{\bm x}}}\right)^2 + 2\ln p(\epsilon^\ast_{\bm x})+\ln(2\pi\sigma^2_{\epsilon^\ast_{\bm x}})
        \;.
    \end{split}
    \end{equation*}
}%
where we have introduced the signal-to-noise ratio (SNR) for template $\bm n^{(i)}$, 
\begin{equation} \label{eq:snr}
    \text{SNR}_i(\bm x) 
    = \frac{\epsilon^\ast_{\bm x}}{\sigma_{\epsilon^\ast_{\bm x}}}
    \stackrel{\text{\ref{eq:epsilon_gauss}, \ref{eq:sigma_gauss}}}{=}  \frac{\Delta \bm x^T \Sigma^{-1} \bm n^{(i)}}{\sqrt{(\bm n^{(i)})^T \Sigma^{-1} \bm n^{(i)}}} \;,
\end{equation}
and the constant 
\begin{equation} \label{eq:c}
    C=2\ln p(\epsilon^\ast_{\bm x})+\ln(2\pi\sigma^2_{\epsilon^\ast_{\bm x}})\;.
\end{equation}

To sum up, in the case of a general Gaussian likelihood function and general distortion components $\bm n^{(i)}$, the individual test statistics $t_i(\bm x)$ measure the strength of the evidence for the presence of the distortion $\bm n^{(i)}$ in the data (Equation~\ref{eq:tSNR}). This is directly analogous to the matched filtering technique used in signal detection, where the data is correlated with a set of template signals to find the one that maximizes the SNR.

Let us now restrict to the case where the distortions correspond to deviations along the standard basis vectors in data space. Specifically, we set the distortion directions to be the unit vectors in the $i\text{th}$ dimension of the data space, \ie\ $\bm n^{(i)} \equiv \bm e_i$. We find that
\begin{equation}
    \text{SNR}^2_i(\bm x) = \left[\Delta\bm x^T \Sigma^{-1} \Delta\bm x\right]_i \stackrel{\text{\ref{eq:deltax}}}{=} \left[(\bm x - \bm \mu)^T \Sigma^{-1} (\bm x - \bm \mu)\right]_i.
\end{equation}

Thus, considering all the possible alternatives of the standard basis and summing over them
\begin{equation} \label{eq:app_chi2}
    t_\text{sum}(\bm x) = \left[(\bm x - \bm \mu)^T \Sigma^{-1} (\bm x - \bm \mu)\right] + \mathrm{const.}
    = \chi^2 + \mathrm{const.}
\end{equation}
where we have recovered the classical Pearson's $\chi^2$ test up to a constant. We discuss in the next section when the conditions under which this constant is effectively independent of $\bm x$.

\subsection{Choice of prior boundaries} \label{subsec:boundaries}

\noindent In our derivations, we have assumed a uniform prior for the distortion amplitude $\epsilon$, specifically
$p(\epsilon) \equiv \mathcal{U}(-b, b)$. The choice of 
the boundaries $b$, hence of $p(\epsilon)$, affects the constant term $C$ in the test statistic (see Equation~\ref{eq:c}).

One way to make this quantitative is to define a maximum SNR for a given distortion, $\mathrm{SNR}_\mathrm{max}$. This is realized by the maximum value of $\epsilon$, \ie\ $b$. Writing this in terms of the variance $\sigma_{\epsilon^\ast_{\bm x}}$, we see that:
\begin{equation}
    b = \text{SNR}_\text{max} \sigma_{\epsilon^\ast_{\bm x}} = \frac{\text{SNR}_\text{max}}
    {\sqrt{(\bm n^{(i)})^T \Sigma^{-1} \bm n^{(i)}}}
    \;.
\end{equation}
Hence, as long as the prior boundaries $b$ are chosen to be sufficiently wide, the constant $C$ in the test statistic becomes effectively independent of $\bm x$. 

Using the SNR training strategy, described in the following Appendix~\ref{app:SNR}, the prior boundaries $b$ can be adaptively learned by the algorithm, as explained in Section~\ref{subsec:adaptive} and shown in Figure~\ref{fig:adaptive}.

\section{SNR-based training strategy} \label{app:SNR}

\noindent We have seen in Section~\ref{subsec:connection} and in Appendix~\ref{app:math} that the test statistic for a given distortion $\bm{n}^{(i)}$ is directly related to the SNR of that distortion in the data (Equation~\ref{eq:tSNR}), where the SNR is given in Equation~\ref{eq:snr}. Thus, these test statistics can be equivalently trained by minimizing a Gaussian negative log-likelihood loss for the MLE of the matched filter $\epsilon_{\phi, i}(\bm x)$ and its variance $\sigma^2_{\phi, i}$ given a distortion $i$
\begin{multline}
    \mathcal{L}^{(i)}_\mathrm{SNR} \left[\epsilon_{i, \phi}(\bm x), \sigma_{i, \phi}^2\right]  = \\ 
   \mathbb{E}_{\bm x, \epsilon \sim p_\mathrm{dist}(\bm x | i, \epsilon) p(\epsilon)} 
   \left[
    \frac{(\epsilon_{i, \phi}({\bm x}) - \epsilon)^2}{\sigma_{i, \phi}^2} + \ln \sigma_{i, \phi}^2
    \right] \;.
\end{multline}
It is then straightforward to compute the SNR and hence the test statistics of interest from Equation~\ref{eq:tSNR} having the estimates $\epsilon_{i, \phi}({\bm x})$ and $\sigma_{i, \phi}^2$
\begin{equation} \label{eq:t_epsilon}
    \hat{t}_i(\bm x) \propto \frac{\epsilon_{i, \phi}(\bm x)}{\sigma_{i, \phi}} \;.
\end{equation}
As discussed in Appendix~\ref{app:BCE} for the classifier-based training strategy, in the interest of efficiency, rather than training individual networks $\epsilon_{i, \phi}$ and $\sigma^2_{i, \phi}$ for each possible localized distortion $i$, we typically train single multi-output networks of the form
\begin{flalign}  \label{eq:nnSNR}
    \boldsymbol{\epsilon}_{\phi} (\bm x): \mathcal{D} \to \mathbb{R}^{N_\text{alt}}
    \;,\\ \label{eq:nnSNRsigma}
    \boldsymbol{\sigma}_{\phi}^{2}: \mathcal{D} \to \mathbb{R}^{N_\text{alt}}
    \;.
\end{flalign}
The total loss is given as a sum over the individual losses,
\begin{equation}
    \mathcal{L}_\mathrm{SNR}\left[\boldsymbol{\epsilon}_{\phi}(\bm x), \boldsymbol{\sigma}_\phi \right]=
    \sum^{N_\text{alt}}_{i=1} \mathcal{L}^{(i)}_\mathrm{SNR} \left[\epsilon_{i, \phi}(\bm x), \sigma_{i, \phi}^2\right] \;,
\end{equation}
where the sum runs over all hypotheses $H_i$.

\section{Comparison of training strategies and analytical check} \label{app:comparison}

\begin{figure*}
    \centering
    \begin{tikzpicture}
    \node at (0cm,0cm) {\includegraphics[height=12cm,clip]{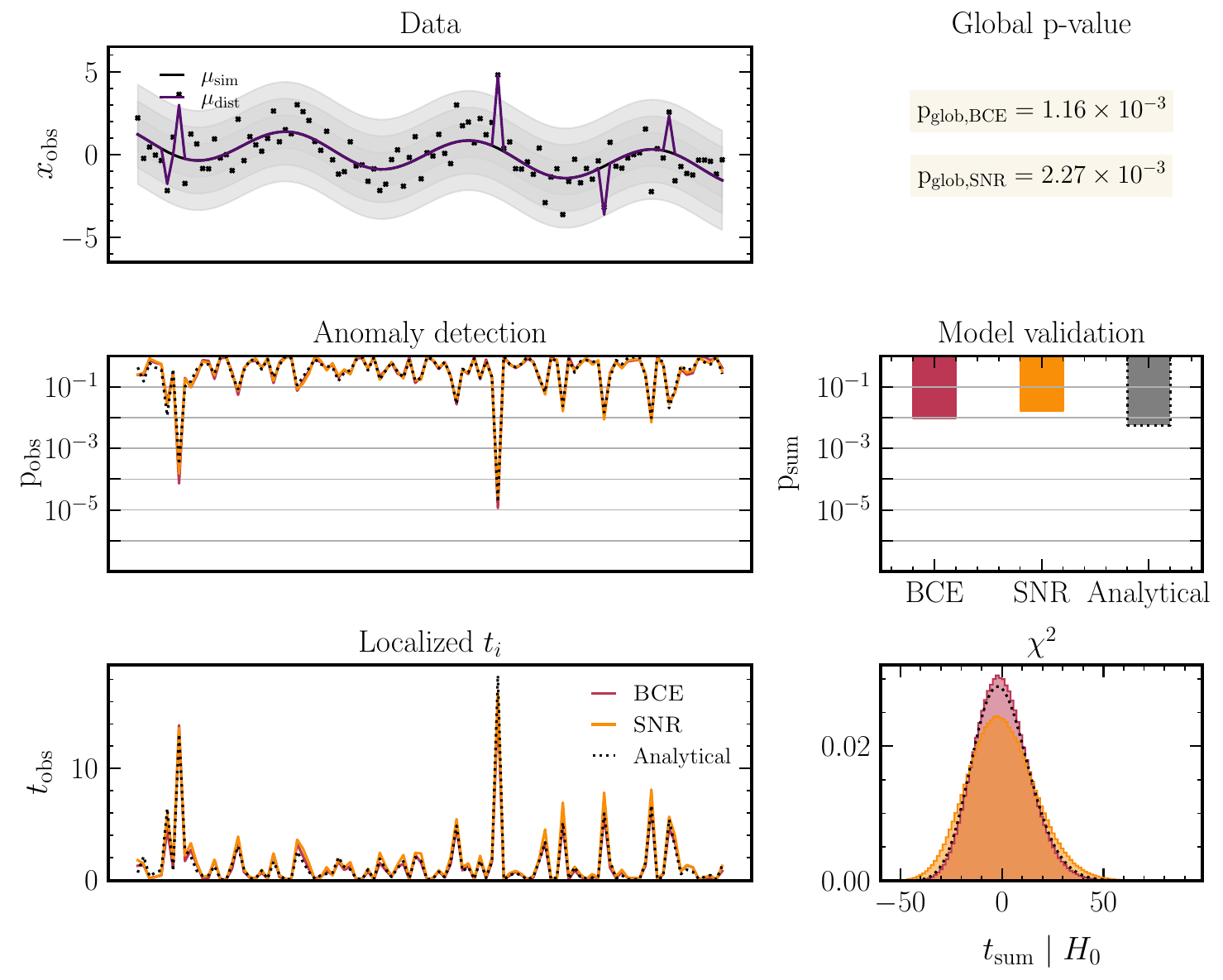}};
    \end{tikzpicture}
    \caption{Similar to Figures~\ref{fig:plot1} and~\ref{fig:plotn}, but for uncorrelated bin-wise distortions. We show in dotted black the analytical (profiled) expectation, in pink the results obtained using the classifier-based (BCE) training strategy presented in Appendix~\ref{app:BCE}, and in orange the results obtained using the SNR-based (SNR) training strategy presented in Appendix~\ref{app:SNR}. The \textbf{upper-left, central-left, and central-right panels} are the same as in Figures~\ref{fig:plot1} and~\ref{fig:plotn}. In the \textbf{lower-left panel} we show the localized test statistics estimated via Equation~\ref{eq:t_class} and  Equation~\ref{eq:t_epsilon} for the pink and orange lines respectively. Finally, in the \textbf{lower-right panel} we show the distribution of the aggregated test statistic $t_\mathrm{sum}$ (Equation~\ref{eq:tsum}) under the null hypothesis (\ie\ applying the network to simulations $\bm x \sim p_\mathrm{sim}(\bm x)$). As discussed in Section~\ref{sec:method} and in Appendix~\ref{app:math}, for simple deviations along the standard basis vector in data space, the aggregated test statistics follows the classical Pearson's $\chi^2$ test.}
    \label{fig:analytical}
\end{figure*}

\begin{figure*}
    \centering
    \begin{tikzpicture}
    \node at (0cm,0cm) {\includegraphics[width=\linewidth,clip]{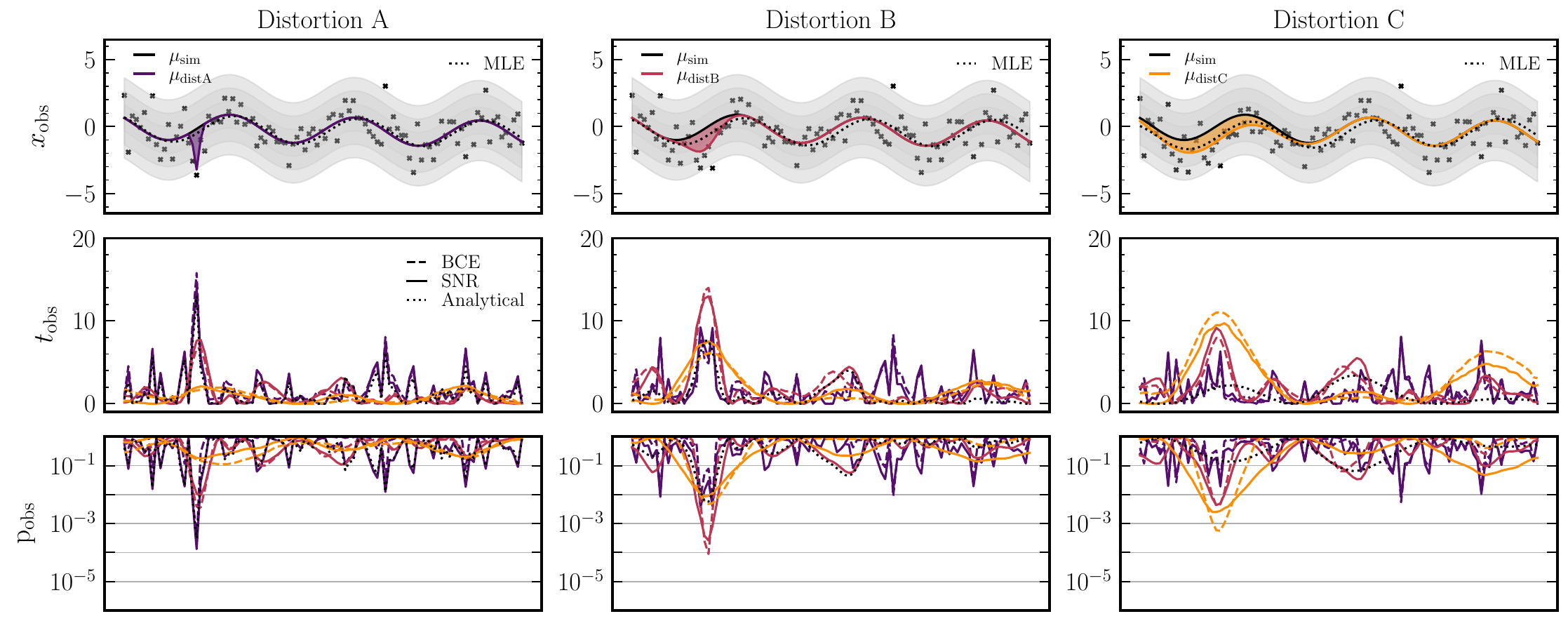}};
    \end{tikzpicture}
    \caption{A comparison of the training strategies and analytical expectation for correlated distortions. Each column is for data distorted by a deviation with different correlation scale, as labeled and color-coded in Figure~\ref{fig:plot1}. The \textbf{first row} shows the data distorted by the distortion, as in the first panel of Figure~\ref{fig:plot1}. We show with a dotted line the MLE prediction. The \textbf{second row} shows the test statistic estimated with the classifier-based training strategy through Equation~\ref{eq:t_class} (dashed lines), with the SNR-based training strategy through Equation~\ref{eq:t_epsilon} (solid lines) or analytically (dotted black line). The \textbf{last row} shows the corresponding significance in terms of $\mathrm{p}$-values. In the last two rows, we can see that the more the MLE prediction absorbs the distortion, the less the neural network-based and the analytical prediction for the test statistic agree. We note in this case that this is the expected behaviour and there is no reason that the two should agree when the MLE is significantly shifted.  
    }
    \label{fig:comparison}
\end{figure*}

\noindent In this section we verify experimentally the connection to classical testing frameworks (Appendix~\ref{app:math}) of our method, and compare the two proposed training strategies (Appendix~\ref{app:BCE} and Appendix~\ref{app:SNR}). Importantly, we derive the analytical quantities for \emph{profiled} likelihoods, while our estimates marginalize over background model variations.

For a first comparison, we further simplify the instructive example presented in Section~\ref{sec:example} by considering only \emph{uncorrelated bin-wise} additive stochastic distortions. In this simple scenario it is straightforward to compute the analytical (profiled) expectation. We show the comparison in Figure~\ref{fig:analytical}. There, we compare the analytical (profiled) expectations (dotted black lines) with the results obtained using our two training strategies --- the classifier-based method from Appendix~\ref{app:BCE} (pink lines) and the SNR-based method from Appendix~\ref{app:SNR} (orange lines) --- in the context of uncorrelated bin-wise additive stochastic distortions. The lower-left panel shows the localized test statistics estimated via Equations~\ref{eq:t_class} (pink lines) and \ref{eq:t_epsilon} (orange lines), where we see that both neural estimators closely match the analytical expectations. As a consequence, also the localized (central-left panel) and aggregated (central-right panel) significances match.
Finally, the lower-right panel presents the distribution of the aggregated test statistic $t_\mathrm{sum}$ under the null hypothesis ($\bm x \sim p_\mathrm{sim}(\bm x)$), confirming that it follows the classical Pearson's $\chi^2$ distribution as discussed in Section~\ref{sec:method} and detailed in Appendix~\ref{app:math}.

We then consider the same setup as in the instructive example presented in Section~\ref{sec:example}, with three correlated distortions of different correlation scales. We show the comparison in Figure~\ref{fig:comparison}. There, we compare the analytical (profiled) expectations (dotted black lines) with the results obtained using our two training strategies. The second row shows the localized test statistics estimated via Equations~\ref{eq:t_class} (dashed lines) and \ref{eq:t_epsilon} (solid lines). In the last two rows, we can see that the two training strategies broadly agree, and in addition that the more the MLE prediction (shown in the first row) absorbs the distortion, the less the neural network-based and the analytical (profiled) prediction for the test statistic agree. We have checked that the analytical predictions and the neural network-based estimates do agree in the absence of a varying background. Thus we conclude that the mismatch is solely due to the MLE compensating the distortion, in case of the analytical profiled test statistic. Note that the neural network-based marginalized test statistics are, by construction, able to pick up the distortion accounting for the model variations.

\end{document}